\documentclass[12pt]{article}
\usepackage{amsmath,amssymb,amsfonts}
\usepackage{cite}

\usepackage{geometry}                
\title{Exact integration of height probabilities \\
in the Abelian Sandpile Model}
\author{
  {Sergio Caracciolo and Andrea Sportiello}
\\[0.3cm]
  {\small\it Dipartimento di Fisica dell'Universit\`a degli Studi di Milano, and INFN,}\\
  {\small\it via Celoria 16, I-20133 Milano, ITALY}\\
  {\small\tt Serg{}io.Caracci{}olo@mi.i{}nfn.it;
\quad An{}drea.Spor{}tiello@mi.in{}fn.it
}}

\newcommand{\be}{\begin{equation}}
\newcommand{\ee}{\end{equation}}

\def\reff#1{(\protect\ref{#1})}

\def\smfrac#1#2{{\textstyle\frac{#1}{#2}}}

\begin{document}

\maketitle

\begin{abstract}
The height probabilities for the recurrent configurations in the
Abelian Sandpile Model on the square lattice have analytic
expressions, in terms of multidimensional quadratures.  At first,
these quantities have been evaluated numerically with high accuracy,
and conjectured to be certain cubic rational-coefficient polynomials
in $\pi^{-1}$. Later their values have been determined by different methods.

We revert to the direct derivation of these probabilities, by computing
analytically the corresponding integrals. Yet another time, we confirm
the predictions on the probabilities, and thus, as a corollary, the
conjecture on the average height, $\langle \rho \rangle = 17/8$.
\end{abstract}

\clearpage

\section{Introduction}

The Abelian Sandpile Model is a non-equilibrium system, driven at a
slow steady rate, with local threshold relaxation rules, which in the
steady state shows relaxation events, called {\em avalanches}, in
bursts of a wide range of sizes and critical spatio-temporal
correlations, obtained without fine-tuning of any control
parameters. We refer to the introductory 
reviews~\cite{Dhar, IP, LP, CPS}.

In the set of {\em stable} configurations in the Abelian Sandpile
Model on (portions of) a square lattice, at each site $i\in
\mathbb{Z}^2$, the height variable can take the values $z_i = 0, 1, 2,
3$\;\footnote{Some authors prefer the values $z_i=1, 2, 3, 4$. Results
  are easily translated between the two notations.}.  Particles are
added randomly and the addition of a particle increases the height at
that site by one. If this height exceeds the critical value $z_c = 3$,
then the site \emph{topples}. On a toppling event, its height
decreases by 4 and the heights at each of its nearest neighbors
increases by~1.

A very natural question is: what is the asymptotic (i.e.,
infinite-volume) probability $P_i$ for the heights $z_i$, for $i= 0,
1, 2, 3$, in the ensemble of {\em recurrent} configurations?

After some numerical studies~\cite{Zhang, Manna, Grassberger},
the first exact result~\cite{MD} concerns with the probability $P_0$
for a site to be empty:
\begin{subequations}
\label{eqs.Pi}
\begin{align}
P_0 &= \frac{2}{\pi^2} - \frac{4}{\pi^3} 
\,.
\intertext{An analytic expression for the other probabilities was
  obtained in~\cite{Prie93, Prie94}:}
P_1 &= \frac{1}{2} - \frac{3}{2 \pi} - \frac{2}{\pi^2} +
\frac{12}{\pi^3} + \frac{I_1}{4}
\,;\\
P_2 &= \frac{1}{4} + \frac{3}{2 \pi} + \frac{1}{\pi^2} -
\frac{12}{\pi^3} - \frac{I_1}{2} - \frac{3 I_2}{32}
\,;\\
P_3 &= \frac{1}{4} - \frac{1}{\pi^2} + \frac{4}{\pi^3} +
\frac{I_1}{4}  + \frac{3 I_2}{32}
\,;
\end{align}
\end{subequations}
where $I_1$ and $I_2$ are expressed as multiple integrals.  These
results had been obtained by using a mapping from the set of recurrent
configurations onto the set of spanning trees covering the lattice.
These trees are rooted (on the boundary of the lattice, where
dissipation occurs). Introduce the concept of \emph{predecessor}:  a vertex $j$
precedes a vertex $i$ if the unique path on the spanning tree from $j$ to the root includes $i$. 
Then, the probabilities $P_k$ at the vertex $i$ are simply related to the numbers $X_k$ of spanning 
trees in which the vertex $i$ has exactly $k$ predecessors among its nearest neighbours. And the 
$X_k$'s can be, at the end, expressed in terms of the lattice Green function.

Furthermore, an indirect argument fixes a relation between $I_1$ and
$I_2$ (see equation (\ref{ipo2}) later on).\footnote{A certain
  combinatorial quantity, know to be finite, is formulated as a
  lattice integral presenting a divergence: tuning to zero an overall
  factor in the divergent part gives the forementioned relation.}
Thus \emph{any} single further linearly-independent information on
$I_1$ and $I_2$, or on the $P_k$'s, would have fixed the height
probabilities completely.

An extensive account on the derivation of these results is provided
in~\cite{Ruelle}, together with several other interesting
properties\,\footnote{In~\cite{Ruelle}, in equation (4.1), the authors
  also correct a misprint in equation (32) of~\cite{Prie94} which was
  wrong by a factor 2. Here we notice a misprint in their equation
  (4.8), where the second term on the right-hand side should be
  $7/\pi$ instead of $7/\pi^2$.}.

As a corollary, the average density in the ensemble of recurrent
configurations is given by
\be
\langle \rho \rangle
=
\sum_{k=0}^3 k\, P_k
=
\frac{7}{4} + \frac{3}{2\pi} - \frac{3}{\pi^2} + \frac{3 I_2}{32}
\,.
\label{eq.46578634}
\ee
As reported in~\cite{Dhar}, this quantity was conjectured by
Grassberger to be\,\footnote{These authors use the range $1 \leq
  z_i \leq 4$, and accordingly write $\langle \rho \rangle  = 25/8$.}
\be
\langle \rho \rangle  = \frac{17}{8}
\,.
\ee
An interesting observation is the following:  the expectation value of
the height probabilities don't help in the understanding of the conformal
features of the corresponding field theory in the continuum, at least in the
whole plane.  But it is not so in the presence of a boundary. Indeed, 
in~\cite{Pirouxa, Pirouxb, RuelleLett, Ruelle}, the evaluation of the height probabilities in the
upper half plane has been used to reveal that the continuum theory is a
logarithmic conformal theory with central charge $c = -2$.
Afterwards, also two-point correlation functions for the height variables have been computed~\cite{Ruelle1005}
and found in agreement with the prediction of a logarithmic conformal field theory based on field identifications 
obtained previously.

A rejuvination of the interest in the exact determination of $\langle
\rho \rangle$ has arised with the work of Fey, Levine and Wilson
\cite{FLW1, FLW2}, in which a subtle difference has been elucidated
between the uniform average on the ensemble of recurrent
configurations, and the properties of the critical system with
conserved mass. As the discrepances in densities between the two
regimes are numerically very small ($\delta\hspace*{-.75pt}\rho / \rho
\sim 10^{-4}$), although the numerical determination of the integral
$I_2$ appearing in (\ref{eq.46578634}) has a much higher precision, it
would have been more satisfactory to have an exact result for at least
one of these two quantities.

In fact, it was possibly in part this rejuvinated interest that led
some time later to two independent proofs, methodologically similar,
of the density conjecture (and, through the argument above, of all the
height probabilities) \cite{RuProof, KWProof}. The single missing
linear relation has been the intensity of the loop-erased random walk
at a first neighbour of the source of the walk, that is
combinatorially related to the density (and turned out to be $5/16$ on
the square lattice).  The role of the loop-erased random walk in the
uniform spanning tree model (and thus in the Abelian Sandpile Model)
should not be surprising, since the works on the subject culminating
in Propp and Wilson exact sampling algorithm~\cite{proppwilsonHowTo}.

In this paper we shall provide a different proof, conceptually simpler
(although, admittedly, theoretically less illuminating): we shall
revert to the original formulation of the problem, and evaluate
exactly the integrals in question. They have the form of two-loop
Feynman integrals in a two-dimensional scalar field theory on the
lattice.  There has been a long-time effort in order to reduce the

evaluation of lattice Feynman integrals at one-loop level, through
simple algebraic methods, both in momentum space~\cite{CMP, BCP}, and
in coordinate space~\cite{LW}.  These methods have found also
important applications in two dimensions, 
respectively~\cite{CP94, CP95} and~\cite{Shin}. In particular,
in~\cite{Claudia} there is an extension to the triangular lattice,
which can be of help to generalize
our procedure to this system (remark that the study of sandpiles on
the triangular and honeycomb lattices, with the aim of height
probabilities and correlation functions, has also been considered
in~\cite{Ruelle0912, KWProof}).

\subsection{The integrals to evaluate}

We shall use the following notations of \emph{lattice momenta}, which
are common in lattice field theory
\begin{align}
\label{eq.lattmom}
\overline{p_\mu} := & \, \sin p_\mu \\
\hat{p}_\mu := & \, 2 \, \sin \frac {p_\mu}{2}
\end{align}
where, in our two-dimensional case, the index $\mu$ can take two
values that we choose to be $0$ and $1$. Then
\be
\hat{p}^2 :=   \, \sum_{\mu = 0,1} \hat{p}_\mu^2 
\ee
is the quantity, invariant under the lattice symmetry, which appears
in the lattice propagator
\be
\Delta(p) :=  \, \frac{1}{\hat{p}^2 + h} \, .
\ee
We have added the regulator of the infrared singularity $h$ just to
have well-defined quantities in all our manipulations, but we are
interested only in the limit of vanishing $h$ (and thus, to integrals
for which this limit exists). In the following we will not mention
explicitly the regulator $h$, and the extraction of the limit will be
understood where pertinent.  Given the shorthands
\begin{align}
\int dp 
&:=  \, \int_{- \pi}^\pi \frac{d p_0}{2 \pi} 
\int_{- \pi}^\pi \frac{d p_1}{2 \pi}\,;
&
\delta_2(p)
&:=
(2 \pi)^2 \delta(p_0)\, \delta(p_1)
\,;
\end{align}
we want to integrate polynomial expressions in the lattice momenta
(\ref{eq.lattmom}), in the measure
\be
d\mu := dp\, dq\, dk\, 
\delta_2(p+q+k)
\;
\Delta(p) \Delta(q) \Delta(k)
\ee
which is invariant under all the permutations of the momenta $p,q,k$,
under exchange of the indices $0$ with $1$, and under simultaneous
inversion of all the momenta along one of the lattice axis. These
invariances imply relations between the integral of different
polynomials, to which we will refer generically in the following as
``symmetry of the integration measure''.
In particular, we use the symbol $A \to B$ to denote the fact 
$\int d\mu \, A = \int d\mu \, B$.

In order to define the integrands pertinent to the expressions in
(\ref{eqs.Pi}), we have to start form the matrix $M(c_1, c_2, c_3)$,
given in \cite[eq.\ (3.18)]{Ruelle}
\be
M(c_1, c_2, c_3) = 
\begin{pmatrix}
c_1 & 1 & e^{i q_1} & 1 \\
c_3 & e^{i p_0 + i q_0}  & e^{- i q_0 + i q_1} & e^{-i p_0} \\
c_2 & e^{i p_1 + i q_1} & 1 & e^{- i p_1} \\
c_2 & e^{-i p_1 - i q_1} & e^{2 i q_1} & e^{i p_1}
\end{pmatrix}
\,.
\label{defgoal1}
\ee
The interesting quantity is the integral \cite[eq.\ (3.17)]{Ruelle}
\be
I(c_1, c_2, c_3) \, = \, \int d\mu \, i \sin p_0 \, 
\det M(c_1, c_2, c_3)\, . 
\label{main}
\ee
It is soon realized that the integral is real, does not depend from
$c_3$, and is, of course, linear in $c_1$ and $c_2$:
\be
I(c_1, c_2, c_3) \, = \, \frac{1}{8} \left(J_1 c_1 + J_2 c_2\right)
\,.
\label{defgoal2}
\ee
The factor $1/8$ is due to our choice to maintain the usual definition
of the lattice propagator. This differs from the choice
in~\cite{Prie93, Prie94, Ruelle} by a factor 2, and we have three propagators in the definition of the integration
 measure.
Then, for the quantities $I_1$, $I_2$ defined above in (\ref{eqs.Pi}),
\begin{align}
I_1 &= J_1 + \left( \frac{4}{\pi} -1 \right) J_2 
\,;
\\
I_2 &= 8\, J_2 -  \frac{16}{\pi} + \frac{4}{\pi^2}
\,.
\end{align}
In~\cite{Ruelle} it is shown, by an indirect compatibility argument,
that the relation
\be
J_1 + J_2 = \frac{2}{\pi} - \frac{4}{\pi^2} 
\label{ipo2}
\ee
must hold.
This is verified numerically to high precision (to order $10^{-12}$).
It is also observed numerically to high precision that
\be
J_2 = \frac{1}{2}
\,.
\label{ipo1}
\ee 
If these relations hold exactly then 
\begin{subequations}
\begin{align}
P_1 &= \frac{1}{4} - \frac{1}{2 \pi} - \frac{3}{\pi^2} +
\frac{12}{\pi^3}
\,; \\
P_2 &= \frac{3}{8} + \frac{1}{\pi} - \frac{12}{\pi^3} 
\,; \\
P_3 &= \frac{3}{8} -  \frac{1}{2 \pi}  + \frac{1}{\pi^2} +
\frac{4}{\pi^3} 
\,;
\end{align}
\end{subequations}
and also the conjecture by Grassberger on the density follows
\be
\sum_{k=0}^3 k \, P_k  \, = \, \frac{17}{8}
\,.
\ee


\subsection{The strategy}

The details of the derivation are given in the following sections. Let
us outline here the general strategy that we adopted all along the
calculation. By exploiting the symmetry of the integration measure, we
try to obtain, in the numerator, a factor which can cancel one of the
propagators (at vanishing regularisation).  Say that we get a
$\hat{p}^2$ in the numerator. Then, we write all other appearences of
$p_{\mu}$'s as $-(q_{\mu}+k_{\mu})$'s.  Thus, possibly through a
trigonometric expansion at the numerator, the remaining integrals are
factorized in independent one-loop integrals in the two other momenta.
Some useful trigonometric identities used at this aim are
\begin{align}
\sum_{\mu = 0, 1} \overline{p_\mu}^2 
&=
\hat{p}^2 - \frac{1}{4} \sum_{\mu = 0, 1} \hat{p}_\mu^4
\,;
\label{eq.usetrig1}
\\
\hat{k}^2_\mu
&=
\hat{q}^2_\mu + \hat{p}^2_\mu -  \frac{1}{2} \, \hat{q}_\mu^2\,
\hat{p}_\mu^2 + 2 \, \overline{q_\mu}\,
\overline{p_\mu} 
\,.
\label{pqbarra}
\end{align}
The latter, which is valid when the sum of the momenta $p, q$ and $k$
vanishes, is sometimes useful also in the inverse form, in which
$\overline{q_\mu}\, \overline{p_\mu}$ is expressed in terms of the
rest.

We shall need the very elementary one-loop integrals:
\begin{align}
\int dq\, \hat{q}_0^2 
&= 2 
\,;
&
\int dq\, \frac{\overline{q_0}}{\hat{q}^2} 
&= 0
\,;
&
\int dq \,\frac{\hat{q}_0^2}{\hat{q}^2} 
&= \frac{1}{2}
\,;
&
\int dq \,\frac{\hat{q}_0^4}{\hat{q}^2} 
&= \frac{4}{\pi}
\,.
\label{semplici} 
\end{align}
From these building blocks, other integrals soon follow, for example
\begin{align}
\int dq \,\frac{\hat{q}_0^2 \hat{q}_1^2}{\hat{q}^2} 
&=
\int dq \, \frac{\hat{q}_0^2 \,( \hat{q}^2 - \hat{q}_0^2)}{\hat{q}^2}
= 2 - \frac{4}{\pi}
\,;
\\
\int dq \,\frac{\overline{q_0}^2}{\hat{q}^2} 
&=
\int dq \, \left( \frac{\hat{q}_0^2}{\hat{q}^2} + \frac{1}{4}
\frac{\hat{q}_0^4}{\hat{q}^2} \right) 
= \frac{1}{2}  - \frac{1}{\pi}
\,;
\\
\intertext{and the slightly more tricky}
\int dq \, \frac{\hat{q}_0^2 \hat{q}_1^4}{\hat{q}^2}  
&=
\frac{1}{2}
\int dq \, \frac{\hat{q}_0^2 \hat{q}_1^2 
(\hat{q}_0^2 + \hat{q}_1^2)}{\hat{q}^2}
=
\frac{1}{2}
\int dq \, \hat{q}_0^2 \hat{q}_1^2 
= 2
\,.
\end{align}
We shall also need
\be
\int dq\, \frac{\overline{q_0}^2 \, \cos q_1 }{\hat{q}^2} =   \, \int
dq \, \left( \frac{\hat{q}_0^2}{\hat{q}^2} - \frac{1}{4}
\frac{\hat{q}_0^4}{\hat{q}^2} - \frac{1}{2} \frac{\hat{q}_0^2
  \hat{q}_1^2}{\hat{q}^2} + \frac{1}{8} \frac{\hat{q}_0^2
  \hat{q}_1^4}{\hat{q}^2}\right) 
=  \, -\frac{1}{4} + \frac{1}{\pi} \, .
\ee
One more trigonometric identity has been used in the Appendix
in order to compute a slightly more complex integral. 


\section{The integral $J_2$}


The contribution proportional to $c_2$ comes from the integral of
\begin{multline}
2 \sin p_0 \sin q_1 \left\{ \sin p_1 [ ( - \cos q_0 + \cos q_1) 
 \sin p_0 - ( 1 + \cos p_0 + \cos q_1) \sin q_0]  \right. \\
\left. + \sin q_1 [( 1 + \cos p_1 + \cos q_0) \sin p_0 + 
 (\cos p_0 - \cos p_1) \sin q_0 ]\right\}
\end{multline}
which has been obtained, from the definitions
(\ref{defgoal1}--\ref{defgoal2}), by performing a symmetrization by
first changing all the signs of $p_0, q_0$, and afterwards of $p_1,
q_1$.
A convenient rewriting, using $k+p+q=0$, is
\be
\label{eq.76354765}
2\, \overline{p_0} \,  \overline{q_1} \, 
\left[ 
(\overline{p_0} \,  \overline{q_1}  - \overline{q_0} \, \overline{p_1})
+ 
(\overline{q_0} \,  \overline{k_1}  - \overline{k_0} \, \overline{q_1})
+
(\overline{k_0} \,  \overline{p_1}  - \overline{p_0} \, \overline{k_1})
\right]
\,,
\ee
that is, through a complete symmetrisation,
\be
\frac{1}{3}\,
\left[ 
(\overline{p_0} \,  \overline{q_1}  - \overline{q_0} \, \overline{p_1})
+
(\overline{q_0} \,  \overline{k_1}  - \overline{k_0} \, \overline{q_1})
+
(\overline{k_0} \,  \overline{p_1}  - \overline{p_0} \, \overline{k_1})
\right]^2
\,.
\ee
The expression in square brackets has a nice geometrical interpretation: it corresponds
to twice the area of the triangle with vertices located at the points
$(\overline{p_0},\,\overline{p_1})$, 
$(\overline{q_0},\,\overline{q_1})$ and
$(\overline{k_0},\,\overline{k_1})$.

A different rewriting, reverting to $p$ and $q$ only, and changing $p$
with $q$ in some terms of (\ref{eq.76354765}), is
\be
\left( \overline{p_0} \,  \overline{q_1}
- \overline{q_0} \, \overline{p_1} \right)^2
+2 \, \overline{p_0} \,
\overline{p_0+ q_0} \, \overline{q_1} \, 
\left( \overline{q_1} - \overline{p_1}\right) 
+ 2 \, \overline{p_1} \, \overline{p_1+ q_1} \,
\overline{q_0} \, \left( \overline{q_0} - \overline{p_0} \right) 
\,.
\label{ris}
\ee
The last two summands clearly give identical result after integration,
and we have to evaluate two integrals
\begin{align}
\label{eq.J2a}
J_2^{(a)}
&=
\int d\mu \,
\left( \overline{p_0} \,  \overline{q_1}
- \overline{q_0} \, \overline{p_1} \right)^2
\,;
\\
\label{eq.J2b}
J_2^{(b)}
&=
\int d\mu \,
4\, \overline{p_0} \,
\overline{p_0+ q_0} \, \overline{q_1} \, 
\left( \overline{q_1} - \overline{p_1}\right) 
\,;
\end{align}
so that $J_2 = J_2^{(a)} + J_2^{(b)}$.

\subsection{The integral $J_2^{(a)}$}


We start by evaluating the integral in (\ref{eq.J2a}).  We rewrite it
as

\be
\bigg[ \bigg( \sum_{\mu = 0, 1} \overline{p_\mu}^2 \bigg) 
\bigg( \sum_{\nu = 0, 1} \overline{q_\nu}^2 \bigg) 
-\bigg( \sum_{\mu = 0, 1} \overline{p_\mu}\, \overline{q_\mu} \bigg)^2 
\,\bigg]
\ee
In the integration, exploiting the symmetries,
\be
\begin{split}
\bigg( \sum_{\mu = 0, 1} \overline{p_\mu}^2 \bigg) 
\bigg( \sum_{\nu = 0, 1} \overline{q_\nu}^2 \bigg) 
&= 
\bigg( \hat{p}^2 - \frac{1}{4} \sum_{\mu = 0, 1} \hat{p}_\mu^4 \bigg) 
\bigg( \hat{q}^2 - \frac{1}{4} \sum_{\nu = 0, 1} \hat{q}_\nu^4 \bigg) \\
& \to
\,\, \hat{p}^2 \hat{q}^2 - \hat{p}^2 \hat{q}_0^4 
+ \frac{1}{16} \bigg( \sum_{\mu = 0, 1} \hat{p}_\mu^4 \bigg) 
\bigg(\sum_{\nu = 0, 1} \hat{q}_\nu^4 \bigg) 
\,.
\label{similar}
\end{split}
\ee
Using \reff{pqbarra} in the inverse form, the expansion of the square
produces 3 contributions. The first one is
\be
\begin{split}
&
-\frac{1}{4} \big( \hat{k}^2 - \hat{q}^2 - \hat{p}^2 \big)^2 
\to \,
-\frac{1}{4} \Big[ 3 
\big(\hat{k}^2\big)^2 - 2\,  \hat{q}^2 \hat{p}^2 \Big] \\
& \qquad = 
-\frac{3}{4} \, \hat{k}^2 
\bigg( \hat{p}^2 + \hat{q}^2 - \frac{1}{2} \sum_{\mu = 0, 1}
\hat{q}_\mu^2\, \hat{p}_\mu^2 
+ 2 \sum_{\mu = 0, 1} \overline{p_\mu}\, \overline{q_\mu} 
\bigg) 
+ \frac{1}{2}\,  \hat{q}^2 \hat{p}^2 \\
& \qquad \to \, 
-\hat{q}^2 \hat{p}^2  + \frac{3}{16} \, \hat{k}^2  \hat{q}^2 \hat{p}^2
\,,
\end{split}
\ee
where we used also the fact that in the integration the factor
$\hat{k}^2$ cancels all the dependence from $k$ and the subsequent
integration of $\overline{p_\mu}\, \overline{q_\mu}$ vanishes, while
the integration of $\hat{q}_\mu^2\, \hat{p}_\nu^2$ does not depend on
the values of $\mu$ and $\nu$.

The second contribution is
\be
\begin{split}
-\frac{1}{4} 
\big( \hat{k}^2 - \hat{q}^2 - \hat{p}^2 \big) \sum_{\mu = 0, 1} \hat{q}_\mu^2\, \hat{p}_\mu^2  \to
& -  \frac{1}{8} \, \hat{k}^2  \hat{q}^2 \hat{p}^2 + \frac{1}{2} \,
\hat{p}^2 \sum_{\mu = 0, 1} \hat{q}_\mu^2\, \widehat{(q+k)}_\mu^2 \\
\to & -  \frac{1}{8} \, \hat{k}^2  \hat{q}^2 \hat{p}^2 + \hat{p}^2
\hat{q}_0^2 \, \big( \hat{q}_0^2 +  \hat{k}_0^2  - \smfrac{1}{2}
\hat{q}_0^2 \hat{k}_0^2\big) \\
\to & -  \frac{1}{8} \, \hat{k}^2  \hat{q}^2 \hat{p}^2 + \hat{p}^2
\hat{q}_0^4 + \frac{1}{4} \, \hat{k}^2  \hat{q}^2 \hat{p}^2  -
\frac{1}{4} \, \hat{p}^2 \hat{k}^2 \hat{q}_0^4\, .
\end{split}
\ee
We now combine the last term with the similar one in~\reff{similar},
that is
\be
\begin{split}
\frac{1}{16} & \bigg[ \bigg( \sum_{\mu = 0, 1} \hat{p}_\mu^4 \bigg)
  \bigg(\sum_{\nu = 0, 1} \hat{q}_\nu^4 \bigg) -  \bigg( \sum_{\mu =
    0, 1} \hat{p}_\mu^2 \hat{q}_\mu^2 \bigg)^2 \bigg] = \,
\frac{1}{16} \, \big(  \hat{p}_0^2 \hat{q}_1^2 -  \hat{p}_1^2
\hat{q}_0^2 \big)^2 \\
&=
\frac{1}{16} \big[  \hat{p}_0^2  \big( \hat{q}^2 -\hat{q}_0^2 \big) -
  \big( \hat{p}^2 -\hat{p}_0^2 \big)  \hat{q}_0^2 \big]^2
=
\frac{1}{16} \big[ \hat{p}_0^2 \hat{q}^2 - \hat{p}^2 \hat{q}_0^2
  \big]^2 \\
&\to \, \frac{1}{8} \, \hat{k}^2 \,\hat{q}_0^2\, \big( \hat{k}^2
\hat{q}_0^2 - \hat{q}^2   \hat{k}_0^2 \big) \\
&\to \, \frac{1}{8} \, \hat{k}^2 \,\hat{q}_0^2\, \big[ \big( \hat{p}^2
  + \hat{q}^2 - \frac{1}{2} \sum_{\mu = 0, 1} \hat{q}_\mu^2\,
  \hat{p}_\mu^2 \big)  \hat{q}_0^2 - \hat{q}^2 \big(  \hat{p}_0^2 +
  \hat{q}_0^2 - \frac{1}{2}  \hat{q}_0^2\, \hat{p}_0^2 \big) \big] \\
&\to \, \frac{1}{8} \, \hat{k}^2  \hat{p}^2  \hat{q}_0^4 -
\frac{1}{16} \, \hat{k}^2 \hat{q}_0^4 \sum_{\mu = 0, 1}
\hat{q}_\mu^2\, \hat{p}_\mu^2 - \frac{1}{16} \, \hat{k}^2 \hat{p}^2
\hat{q}^2 \hat{q}_0^2 + \frac{1}{16} \, \hat{k}^2 \hat{q}^2
\hat{q}_0^4\, \hat{p}_0^2 \\
& \to \, \frac{1}{8} \, \hat{k}^2  \hat{p}^2  \hat{q}_0^4  -
\frac{1}{16} \, \hat{k}^2 \hat{p}^2  \hat{q}^2 \hat{q}_0^2 
\,.
\end{split}
\ee
By collecting all the pieces, and using the elementary integrals
\reff{semplici}, we get
\be
 \left( \frac{3}{16} +  \frac{1}{8} - \frac{1}{16} \,\hat{q}_0^2
 \right) \hat{k}^2  \hat{q}^2 \hat{p}^2  + \left(\frac{1}{8} -
 \frac{1}{4} \right) \, \hat{k}^2  \hat{p}^2  \hat{q}_0^4  \to
 \left(\frac{3}{16} - \frac{1}{2 \pi}\right) \,\hat{k}^2  \hat{q}^2
 \hat{p}^2
\,.
\ee
In conclusion, we find that the value of the first term is
\be
J_2^{(a)} = \int d\mu \,
\left( \overline{p_0} \, \overline{q_1} - \overline{q_0} \, \overline{p_1} \right)^2 =
 \frac{3}{16} - \frac{1}{2 \pi} \,.
\label{former}
\ee

\subsection{The integral $J_2^{(b)}$}

We now consider the evaluation of the integral in (\ref{eq.J2b}), and
remark that the integrand can be written as
\be
4 \, \overline{p_0} \,
\overline{p_0+ q_0} \, \overline{q_1} \, 
\left( \overline{q_1} - \overline{p_1}\right) 
\to
-2\, \sum_{\mu = 0, 1} \overline{p}_\mu \, \overline{k}_\mu \,
\sum_{\nu = 0, 1} \overline{q}_\nu \,
( \overline{q}_\nu - \overline{p}_\nu ) 
\ee
because the terms with $\mu = \nu$ give a vanishing contribution
(they are anti-symmetric under the exchange of $p$ with
$q$). Repeated use of (\ref{eq.usetrig1}) and (\ref{pqbarra}) gives
\be
 \frac{1}{2}\, \bigg[ (\hat{q}^2 - \hat{k}^2 - \hat{p}^2 ) +
   \frac{1}{2} \sum_{\mu = 0, 1} \hat{p}_\mu^2 \, \hat{k}_\mu^2 \bigg]
 \,
\bigg[(\hat{k}^2 - 3\, \hat{q}^2 - \hat{p}^2 ) +  \frac{1}{2}
  \sum_{\nu = 0, 1} \hat{q}_\nu^4 + \frac{1}{2} \sum_{\nu = 0, 1}
  \hat{p}_\nu^2 \, \hat{q}_\nu^2 \bigg]  \, .
\ee
We split this evaluation into three terms (the following
(\ref{eq.tt1}), (\ref{third}) and (\ref{eq.tt3})).
A first contribution is
\be
\label{eq.tt1}
\begin{split}
&
\frac{1}{2}\, \bigg[ (\hat{q}^2 - \hat{k}^2 - \hat{p}^2 ) +
  \frac{1}{2} \sum_{\mu = 0, 1} \hat{p}_\mu^2 \, \hat{k}_\mu^2 \bigg]
\, (\hat{k}^2 - 3\, \hat{q}^2 - \hat{p}^2 ) \\ 
& \qquad \to \, 
-\frac{3}{2}\, \bigg[ (\hat{q}^2 - \hat{k}^2 - \hat{p}^2 )
+\frac{1}{2} \sum_{\mu = 0, 1} \hat{p}_\mu^2 \, \hat{k}_\mu^2 \bigg]
\, \hat{q}^2
= -3 \, \hat{q}^2 \, \sum_{\mu = 0, 1} \overline{p}_\mu \,
\overline{k}_\mu
\to \, 0
\end{split}
\ee
which vanishes in the integral.
A second contribution is
\begin{align}
\frac{\hat{q}^2 - \hat{k}^2 - \hat{p}^2}{4}
\bigg( \sum_{\nu = 0, 1} \hat{q}_\nu^4 + \sum_{\nu = 0, 1}
\hat{p}_\nu^2 \, \hat{q}_\nu^2 \bigg)
& \to \,
\frac{\hat{q}^2  - 2 \,\hat{k}^2}{4}
\,\sum_{\nu = 0, 1}
\hat{q}_\nu^4 \, -
\frac{\hat{k}^2}{4}\, 
\sum_{\nu = 0, 1} \hat{p}_\nu^2 \, \hat{q}_\nu^2 
\label{third}
\end{align}
We have now, in all summands, an exposed propagator. Following our
general strategy, we rewrite the remaining expressions using
$p+q+k=0$, namely
\be
\begin{split}
 \frac{1}{4}\, \hat{q}^2 \,  \sum_{\nu = 0, 1} \hat{q}_\nu^4
&=
\frac{1}{4}\, \hat{q}^2 \,  \sum_{\nu = 0, 1}  \Big[
   \hat{k}_\nu^2 \Big(1 - \smfrac{1}{4} \hat{p}_\nu^2 \Big) +
   \hat{p}_\nu^2 \Big(1 - \smfrac{1}{4} \hat{k}_\nu^2 \Big) + 2
   \overline{k}_\nu \overline{p}_\nu \Big]^2 \\
& \to \,
\frac{1}{2}\, \hat{q}^2 \,  \sum_{\nu = 0, 1}  \Big[
   \hat{k}_\nu^4 \Big(1 - \smfrac{1}{4} \hat{p}_\nu^2 \Big)^2 + 3\,
   \hat{p}_\nu^2 \Big(1 - \smfrac{1}{4} \hat{p}_\nu^2 \Big)
   \hat{k}_\nu^2 \Big(1 - \smfrac{1}{4} \hat{k}_\nu^2 \Big)\Big] \\
& \to \,
\frac{1}{2}\, \hat{q}^2 \,  \sum_{\nu = 0, 1} \Big(
  \hat{k}_\nu^4 + 3\, \hat{k}_\nu^2 \hat{p}_\nu^2 - 2\,\hat{k}_\nu^4
  \hat{p}_\nu^2 + \smfrac{1}{4} \, \hat{k}_\nu^4 \hat{p}_\nu^4 \Big)
\end{split}
\ee
so that the whole contribution from~\reff{third} is
\be
\hat{q}^2 \,  \sum_{\nu = 0, 1} \left(  \frac{5}{4}\, \hat{k}_\nu^2
\hat{p}_\nu^2 - \,\hat{k}_\nu^4 \hat{p}_\nu^2 + \frac{1}{8} \,
\hat{k}_\nu^4 \hat{p}_\nu^4 \right)
\to  \,  \frac{5}{8}\, \hat{q}^2  \hat{k}^2 \hat{p}^2  - \hat{q}^2
\hat{p}^2 \hat{k}_0^4 + \frac{1}{4}\, \hat{q}^2  \hat{k}_0^4
\hat{p}_0^4 \\
\ee
We are left with the third term
\be
\label{eq.tt3}
\frac{1}{8}\,  \sum_{\mu = 0, 1} \hat{p}_\mu^2 \, \hat{k}_\mu^2 \,
\sum_{\nu = 0, 1} \hat{q}_\nu^2 \,  ( \hat{q}_\nu^2  + \hat{p}_\nu^2 )  
\ee
One summand gives
\be
\frac{1}{8}\,  \sum_{\mu = 0, 1} \hat{p}_\mu^2 \, \hat{k}_\mu^2 \, \sum_{\nu = 0, 1} \hat{q}_\nu^4  \to
\frac{1}{4}\, \hat{p}_0^2 \, \hat{k}_0^2 \, \left( \hat{q}_0^4 +
\hat{q}_1^4 \right) \to \frac{1}{2}\, \hat{p}_0^2 \, \hat{k}_0^2 \,
\hat{q}_0^4 
\ee
because
\be
\begin{split}
\int d\mu \, \hat{p}_0^2 \, \hat{k}_0^2 \, \hat{q}_1^4  
&= \int d\mu \, \hat{p}_0^2 \, \hat{k}_0^2 \, \left( \hat{q}^2 -
\hat{q}_0^2 \right)^2
= \int d\mu \, \hat{p}_0^2 \, \hat{k}_0^2 \, \left[ \hat{q}_0^4 - 2
  \hat{q}_0^2 \hat{q}^2 + \left(\hat{q}^2\right)^2\right] \\
&= \int d\mu \, \hat{p}_0^2 \, \hat{k}_0^2 \, \left[ \hat{q}_0^4  +
  \hat{q}^2  \left(\hat{q}_1^2 - \hat{q}_0^2\right)\right]
= \int d\mu \, \hat{p}_0^2 \, \hat{k}_0^2 \, \hat{q}_0^4
\end{split}
\ee
where we used the fact that
\be
\begin{split}
\int d\mu \, \hat{p}_0^2 \, \hat{k}_0^2 \,  \hat{q}^2
\left(\hat{q}_1^2 - \hat{q}_0^2\right) 
&= \int dp\, dk \, \hat{p}_0^2 \, \hat{k}_0^2 \, 
\Big( 2\,
\hat{p}_1^2 - \smfrac{1}{2} \, \hat{p}_1^2 \hat{k}_1^2 - 2\,
\hat{p}_0^2 + \smfrac{1}{2} \, \hat{p}_0^2 \hat{k}_0^2 \Big) 
\Delta(p)\Delta(k)
\\
&= 2 - \frac{4}{\pi} - \frac{1}{2} \, \left( 2 - \frac{4}{\pi}
\right)^2 - \frac{4}{\pi} + \frac{8}{\pi^2}
= 0
\,.
\end{split}
\ee
The second summand is
\begin{align}
\frac{1}{8}\, \sum_{\mu = 0, 1} \hat{p}_\mu^2 \, \hat{k}_\mu^2 \,
\sum_{\nu = 0, 1} \hat{q}_\nu^2 \, \hat{p}_\nu^2 
\to \frac{1}{4}\, \hat{p}_0^4 \, \hat{k}_0^2 \, \hat{q}_0^2 +
\frac{1}{4}\, \hat{p}_0^2 \, \hat{k}_0^2 \, \hat{p}_1^2 \, \hat{q}_1^2
\end{align}
and
\be
\frac{1}{4}\, \hat{p}_0^2 \, \hat{k}_0^2 \, \hat{p}_1^2 \, \hat{q}_1^2
\to \frac{1}{4}\, \hat{p}_0^2 \, \hat{k}_0^2 \, \hat{p}_1^2 \, 
\left( \hat{q}^2 - \hat{q}_0^2 \right)
= \, 
\frac{1}{8}\, \left( 2 - \frac{4}{\pi} \right) 
- \frac{1}{4}\, \hat{p}_0^2 \, \hat{p}_1^2 \, \hat{k}_0^2 \, \hat{q}_0^2
\ee
while
\be
- \frac{1}{4}\, \hat{p}_0^2 \, \hat{p}_1^2 \, \hat{k}_0^2 \, \hat{q}_0^2 
\to \, 
\frac{1}{4}\, \hat{p}_0^4 \, \hat{k}_0^2 \, \hat{q}_0^2 
- \frac{1}{4}\, \hat{p}^2 \, \hat{p}_0^2 \, \hat{k}_0^2 \, \hat{q}_0^2 \\
\ee
and
\be
- \frac{1}{4}\, \int d\mu\, \hat{p}^2 \, \hat{p}_0^2 \, \hat{k}_0^2 \, \hat{q}_0^2 
= \, - \frac{1}{4}\, \int dk\, dq\, \left[ 2 \hat{k}_0^2 
- \frac{1}{2} \hat{k}_0^2 \hat{q}_0^2 \right] \, \hat{k}_0^2 \, \hat{q}_0^2 
= \, - \frac{1}{\pi} + \frac{2}{\pi^2}
\ee
so that
\be
\frac{1}{8}\, \sum_{\mu = 0, 1} \hat{p}_\mu^2 \, \hat{k}_\mu^2 \, 
\sum_{\nu = 0, 1} \hat{q}_\nu^2 \, \hat{p}_\nu^2 \to 
\frac{1}{2}\, \hat{p}_0^4 \, \hat{k}_0^2 \, \hat{q}_0^2 + \frac{1}{4} 
- \frac{3}{2 \pi} + \frac{2}{\pi^2} \, . 
\ee
By collecting all the pieces
\be
\frac{1}{8}\, \int d\mu\, \sum_{\mu = 0, 1} \hat{p}_\mu^2 \,
\hat{k}_\mu^2 \, \sum_{\nu = 0, 1} \hat{q}_\nu^2 \, ( \hat{q}_\nu^2 +
\hat{p}_\nu^2 ) 
= \int d\mu\, \hat{p}_0^4 \, \hat{k}_0^2 \, \hat{q}_0^2 + \frac{1}{4}
- \frac{3}{2 \pi} + \frac{2}{\pi^2}
\ee
and using the result~\reff{altro}, computed in the appendix, the whole
expression $J_2^{(b)}$ is
\be
\begin{split}
&
-2\,\int d\mu\, \sum_{\mu = 0, 1} \overline{p}_\mu \,
\overline{k}_\mu \, \sum_{\nu = 0, 1} \overline{q}_\nu \, (
\overline{q}_\nu - \overline{p}_\nu ) \\
& \qquad
= \int d\mu\, \left( \hat{p}_0^4 \, \hat{k}_0^2 \, \hat{q}_0^2 +
\frac{5}{8}\, \hat{q}^2 \hat{k}^2 \hat{p}^2 - \hat{q}^2 \hat{p}^2
\hat{k}_0^4 + \frac{1}{4}\, \hat{q}^2 \hat{k}_0^4 \hat{p}_0^4 \right)
+ \frac{1}{4} - \frac{3}{2 \pi} + \frac{2}{\pi^2} \\
& \qquad = 
\left( -1 + \frac{6}{\pi} - \frac{6}{\pi^2} \right) + \frac{5}{8}
- \frac{4}{\pi} + \frac{4}{\pi^2} + \frac{1}{4} - \frac{3}{2 \pi} +
\frac{2}{\pi^2}
= 
-\frac{1}{8} + \frac{1}{2 \pi} 
\,. 
\label{latter}
\end{split}
\ee 
In conclusion, by adding the first and the second result, computed
respectively in \reff{former} and~\reff{latter}, we get
\be
\frac{J_2}{8} = \frac{3}{16} - \frac{1}{2 \pi} - \frac{1}{8} +
\frac{1}{2 \pi} = \frac{1}{16}
\ee
in agreement with the prediction $J_2 = 1/2$.

\section{The integral $J_1$}

As anticipated in the introduction (see equation (\ref{ipo2})), 
it is expected (by an indirect argument) that
\be
\frac{J_1 + J_2}{8}
= \frac{1}{4\pi} - \frac{1}{2 \pi^2} 
\,.
\ee
Similarly to our evaluation of $J_2$ (but, as we will see, in a
simpler way), we can attack directly the evaluation of $J_1+J_2$, and
produce an independent check of the relation above.  Recall that $J_1$
is the contribution to (\ref{main}) proportional to $c_1$, namely
\be
\frac{J_1}{8}
=
\int d\mu \,
2\, \overline{p_0}\, \overline{q_1} \, ( \overline{k_0} \,
\overline{q_1-p_1} + \overline{q_0}\, \overline{p_1-k_1} +
\overline{p_0}\, \overline{k_1-q_1} )
\,.
\ee
Writing $\overline{p-q}=\overline{p} \cos q - \overline{q} \cos p$,
restate the integrand above as
\be
2\, \overline{p_0}\, \overline{q_1} \, \left[ ( \overline{q_0}\,
\overline{p_1} -\overline{p_0}\, \overline{q_1} ) \, \cos k_1 + (
\overline{p_0}\, \overline{k_1} - \overline{k_0}\, \overline{p_1}) \,
\cos q_1 + ( \overline{k_0}\, \overline{q_1} - \overline{k_1}\,
\overline{q_0}) \, \cos p_1 \right]
\ee
and remark that, by the replacing
\be
\cos \theta = 1 - \frac{1}{2}\, \hat{\theta}^2
\ee
all the contributions in which we take the $1$, that is
\be
2 \overline{p_0}\, \overline{q_1} \, 
\left[ ( \overline{q_0}\, \overline{p_1} -\overline{p_0}\, \overline{q_1} ) 
+ ( \overline{p_0}\, \overline{k_1} - \overline{k_0}\, \overline{p_1}) 
+ ( \overline{k_0}\, \overline{q_1} - \overline{k_1}\, \overline{q_0}) \right]
\ee
are exactly $-J_2/8$, thus, if we keep only the other terms, we have
\be
\frac{J_1+J_2}{8}
=
-
\int d\mu \,
\overline{p_0}\, \overline{q_1} \, \left[ ( \overline{q_0}\,
  \overline{p_1} -\overline{p_0}\, \overline{q_1} ) \, \hat{k}_1^2 + (
  \overline{p_0}\, \overline{k_1} - \overline{k_0}\, \overline{p_1})
  \, \hat{ q}_1^2 + ( \overline{k_0}\, \overline{q_1} -
  \overline{k_1}\, \overline{q_0}) \, \hat{ p}_1^2 \right]
\,.
\ee
Manipulate the integrand by exchanging $q$ with $p$, and index $0$
with $1$, to get
\be
-\overline{p_0}\, \overline{q_1} \, \left[ ( \overline{q_0}\,
  \overline{p_1} -\overline{p_0}\, \overline{q_1} ) \, \frac{1}{2}\,
  \hat{k}^2 + ( \overline{p_0}\, \overline{k_1} - \overline{k_0}\,
  \overline{p_1}) \, \hat{ q}^2 \right] \to \overline{p_0}^2\,
\overline{q_1}^2 \, \frac{1}{2}\, \hat{k}^2 + \overline{p_0}^2\, \cos
p_1\, \overline{k_1}^2 \, \hat{ q}^2 \, .
\ee
In conclusion
\be
\frac{1}{8} \, (J_1 + J_2) = \frac{1}{2} \left( \frac{1}{2} -
\frac{1}{\pi} \right)^2 + \left( \frac{1}{2} - \frac{1}{\pi} \right)
\left(-\frac{1}{4} + \frac{1}{\pi} \right) =
\frac{1}{2 \pi} \left( \frac{1}{2} - \frac{1}{\pi} \right)
\ee
as it was predicted.

\section{Conclusion}

We have been able to analytically compute some lattice integrals that,
through the work of \cite{Prie93,Prie94,RuelleLett,Ruelle} and
references therein, describe the height probabilities in the ensemble
of recurrent configurations of the Abelian Sandpile Model on the
square lattice, in the thermodynamic limit.

The numerical values of these integrals were already known with high
precision, and the exact expressions solidly conjectured, as
rational-coefficient polynomials in $\pi^{-1}$. Most importantly, a
recent indirect calculation of statistical properties of the loop
erased random walk, or equivalently of domino tilings with prescribed
local patterns of monomers and dimers, was sufficient to determine
completely these values~\cite{RuProof, KWProof}.

Nonetheless, our direct evaluation of the original lattice integrals,
with their strikingly simple results, could be of some interest, and
of some use for future work in similar contexts.

Let us stress again that this result is not based on any new deeper
understanding of the properties of the sandpile model, but completely
relies on elementary trigonometry and symmetry considerations, mainly
with the aim of reducing two-loop lattice integrals to quadratic
polynomials in one-loop integrals.
In particular, at some point we used results previously obtained in
\cite{CP94}. In principle, we do not see any obstacle to recover
similar results on other two-dimensional regular lattices.

\appendix

\section{One more integral}
\label{app}

We need the evaluation of the integral
\be
\int d\mu \, \hat{k}_0^4 \hat{p}_0^2 \hat{q}_0^2 \, .
\ee
We first observe that
\be
(\overline{p_0}\, \overline{q_0}) \, \overline{k_0}^2 
= 
\left[ \frac{1}{2} ( \hat{k}_0^2 - \hat{p}_0^2 - \hat{q}_0^2) 
+ \frac{1}{4} \hat{p}_0^2 \hat{q}_0^2 \right] \left( \hat{k}_0^2 -
\frac{1}{4} \hat{k}_0^4 \right) 
\,,
\label{alpha}
\ee
but it is also
\be
(\overline{p_0}\, \overline{k_0}) \, (\overline{q_0} \,
\overline{k_0}) 
= 
\left[ \frac{1}{2} ( \hat{q}_0^2 - \hat{p}_0^2 - \hat{k}_0^2) +
  \frac{1}{4} \hat{p}_0^2 \hat{k}_0^2 \right] \,
\left[ \frac{1}{2} ( \hat{p}_0^2 - \hat{q}_0^2 - \hat{k}_0^2) +
  \frac{1}{4} \hat{q}_0^2 \hat{k}_0^2 \right]
\,.
\label{beta}
\ee
By difference of~\reff{alpha} and~\reff{beta} we get the trigonometric
identity
\begin{multline}
\hat{k}_0^4 \hat{p}_0^2 \hat{q}_0^2 = 2\, (\hat{p}_0^4 + \hat{q}_0^4 +
\hat{k}_0^4) - 4\, (\hat{p}_0^2 \hat{k}_0^2 + \hat{k}_0^2 \hat{q}_0^2
+ \hat{q}_0^2 \hat{p}_0^2) + 4\, \hat{p}_0^2 \hat{q}_0^2 \hat{k}_0^2
\\ 
+ 2\, \hat{k}_0^4 ( \hat{p}_0^2 + \hat{q}_0^2) - \hat{k}_0^2 (
\hat{p}_0^4 + \hat{q}_0^4) - \hat{k}_0^6
\end{multline}
so that
\be
\hat{k}_0^4 \hat{p}_0^2 \hat{q}_0^2 
\to
6\, \hat{p}_0^4 - 12\, \hat{p}_0^2 \hat{q}_0^2 + 4\,
\hat{p}_0^2 \hat{q}_0^2 \hat{k}_0^2 + 2\, \hat{p}_0^4 \hat{q}_0^2 -
\hat{p}_0^6
\,.
\ee
Let us start with 
\be
6\, \int d\mu \, \left(\hat{p}_0^4 - 2\, \hat{p}_0^2
\hat{q}_0^2\right) 
= - 12 \, \int d\mu \, \left( 2 \, \overline{p_0}^2 + 4\,
\overline{p_0} \, \overline{q_0}\right) 
= - 12\, G_2 = - \frac{1}{4} 
\,;
\ee
where the integral $G_2$ was defined and calculated numerically
in~\cite{Falcioni-Treves} and subsequently computed
in~\cite[eq.\;(A.9)]{CP94}. Then,
\be
4\, \int d\mu \, \hat{p}_0^2 \hat{q}_0^2 \hat{k}_0^2 = 4 A^{(3)} =
\frac{1}{2}
\,;
\ee
where the integral $A^{(3)}$ had been introduced and computed
in~\cite[eq.\;(A.6)]{CP94}. For the evaluation of the last term we use the
same trick that was used in \cite{CP94} to compute $A^{(3)}$, that is
use the fact that
\be
\begin{split}
\int d\mu \,\hat{p}_0^4 \, \left( 2\, \hat{q}_0^2 - \hat{p}_0^2
\right)
&=
\int d\mu \, \left( 2\, \hat{p}_1^4 \hat{q}_1^2 -
\hat{p}_1^6 \right) \\
&=
\int d\mu \, (\hat{p}^2 - \hat{p}_0^2)^2 \left[ 2\, \hat{q}^2 -
  \hat{p}^2 - (2\, \hat{q}_0^2 - \hat{p}_0^2) \right]
\,;
\end{split}
\ee
which has the consequence that
\be
\begin{split}
\int d\mu \,\hat{p}_0^4 \, \left( 2\, \hat{q}_0^2 - \hat{p}_0^2
\right)
&=
\frac{1}{2} \, \int d\mu \, \left[ \hat{p}_1^4 ( 2\, \hat{q}^2 -
  \hat{p}^2) - \hat{p}^2 ( \hat{p}_1^2 - \hat{p}_0^2) ( 2\,
  \hat{q}_0^2 - \hat{p}_0^2) \right] \\
&=
\frac{1}{2} \, \int d\mu \, \hat{p}^2 \, \left[ 2\, \hat{q}_1^4 -
  \hat{p}_1^4 - ( \hat{p}_1^2 - \hat{p}_0^2) ( 2\, \hat{q}_0^2 -
  \hat{p}_0^2) \right] 
\,;
\end{split}
\ee
so that
we are left only with elementary evaluations, that bring us to
\be
\int d\mu \,\hat{p}_0^4 \, \left( 2\, \hat{q}_0^2 - \hat{p}_0^2
\right) = -\frac{5}{4} + \frac{6}{\pi} - \frac{6}{\pi^2} 
\,.
\ee
In conclusion
\be
\int d\mu \, \hat{k}_0^4 \hat{p}_0^2 \hat{q}_0^2 
=
-\frac{1}{4} + \frac{1}{2} -\frac{5}{4} + \frac{6}{\pi} -
\frac{6}{\pi^2} 
= -1 + \frac{6}{\pi} - \frac{6}{\pi^2}
\,.
\label{altro}
\ee

\vfill\eject 



\begin{thebibliography}{99}

\bibitem{Dhar}
D.~Dhar,
{\it Theoretical studies of self-organized criticality,}
Physica A{\bf 369} (2006) 29.

\bibitem{IP}
E.~V.~Ivashkevich and V.~B.~Priezzhev,
{\it Introduction to the sandpile model,}
Physica A{\bf 254} (1998) 97.

\bibitem{LP}
L.~Levine and J.~Propp, 
{\it What is a sandpile?,}
Notices of the AMS {\bf 57}
(2010) 976.

\bibitem{CPS}
S.~Caracciolo, G.~Paoletti and A.~Sportiello, 
{\it Multiple and inverse topplings in the Abelian Sandpile Model,}
to be published on Eur.\ Phys.\ J.\ ST, \\
{\tt arXiv:1112.3491}

\bibitem{Zhang} 
Y.~C.~Zhang, 
{\it Scaling Theory of Self-Organized Criticality,}
Phys.\ Rev.\ Lett.\ {\bf 63} (1989) 470.


\bibitem{Manna}
S.~S.~Manna,
{\it Large-scale simulation of avalanche cluster distribution in 
 sand pile model,}
J.\ Stat.\ Phys.\ {\bf 59} (1990) 509.

\bibitem{Grassberger}
P.~Grassberger and S.~S.~Manna, 
{\it Some more sandpiles,}
J.\ Phys.\ (Paris) {\bf 51} (1990) 1077.

\bibitem{MD}
S.~N.~Majumdar and D.~Dhar, 
{\it Height correlations in the Abelian sandpile model,}
J.\ Phys.\ A {\bf 24} (1991) L357.

\bibitem{Prie93}
V.~B.~Priezzhev, 
{\it Exact Height Probabilities in the Abelian Sandpile Model,}
Physica Scripta T{\bf 49} (1993) 663.

\bibitem{Prie94}
V.~B.~Priezzhev, 
{\it Structure of two-dimensional sandpile. I. Height probabilities,}
J.\ Stat.\ Phys.\ {\bf 74} (1994) 955.

\bibitem{Pirouxa}
G.~Piroux and P.~Ruelle,
{\it Pre-logarithmic and logarithmic fields in a sandpile model,}
J.\ Stat.\ Mech.\ (2004) P10005, 
{\tt arXiv:hep-th/0407143}.

\bibitem{Pirouxb}
G.~Piroux and P.~Ruelle,
{\it Boundary height fields in the Abelian sandpile model,}
J.\ Phys.\  A {\bf 38} (2005) 1451,
{\tt arXiv:hep-th/0409126}.


\bibitem{RuelleLett}
G.~Piroux and P.~Ruelle,
{\it Logarithmic scaling for height variables in the Abelian sandpile
  model,}
Phys.\ Lett.\ B{\bf 607} (2005) 188, 
{\tt arXiv:cond-mat/0410253}.

\bibitem{Ruelle}
M.~Jeng, G.~Piroux and P.~Ruelle,
{\it Height variables in the Abelian sandpile model: scaling fields and correlations,}
J.\ Stat.\ Mech.\ (2006) P10015,
{\tt arXiv:cond-mat/0609284}.

\bibitem{Ruelle1005}
V.~S.~Poghosyan, S.~Y.~Grigorev, V.~B.~Priezzhev and P.~Ruelle,
{\it Logarithmic two-point correlators in the Abelian sandpile model,}
J.\ Stat.\ Mech.\ 
(2010) P07025,
{\tt arXiv:1005.2088}.

\bibitem{FLW1}
A.~Fey, L.~Levine and D.~B.~Wilson,
{\it Driving sandpiles to criticality and beyond,}
Phys.\ Rev.\ Lett.\ {\bf 104} (2010) 145703, 
{\tt arXiv:0912.3206}.

\bibitem{FLW2}
A.~Fey, L.~Levine and D.~B.~Wilson,
{\it The approach to criticality in sandpiles,}
Phys.\ Rev.\ E {\bf 82} (2010) 031121,
{\tt arXiv:1001.3401}.

\bibitem{RuProof}
V.~S.~Poghosyan, V.~B.~Priezzhev and P.~Ruelle,
{\it Return probability for the loop-erased random walk and mean
  height in sandpile: a proof},
J.\ Stat.\ Mech.\ (2011) P10004,
{\tt  arXiv:1106.5453}.

\bibitem{KWProof}
R.~W.~Kenyon and D.~B.~Wilson,
{\it Spanning trees of graphs on surfaces and the intensity of
  loop-erased random walk on $\mathbb{Z}^2$,}
{\tt  arXiv:1107.3377}.





\bibitem{proppwilsonHowTo}
J.~G.~Propp and D.~B.~Wilson,
{\it How to get a perfectly random sample from a generic Markov chain
  and generate a random spanning tree of a directed graph,}
Journal of Algorithms {\bf 27} (1998) 170. 

\bibitem{CMP}
S.~Caracciolo, P.~Menotti and A.~Pelissetto, 
{\it One loop analytic computation of the energy momentum tensor for
 lattice gauge theories,}
Nucl.\ Phys.\ B{\bf 375} (1992) 195. 

\bibitem{BCP}
G.~Burgio, S.~Caracciolo and A.~Pelissetto, 
{\it Algebraic algorithm for the computation of one-loop Feynman
 diagrams in lattice QCD with Wilson fermions,}
Nucl.\ Phys.\ B{\bf 478} (1996) 687. 

\bibitem{LW}
M.~L\"{u}scher and P.~Weisz,
{\it Coordinate space methods for the evaluation of Feynman diagrams
 in lattice field-theories,}
Nucl.\ Phys.\ B{\bf 445} (1995) 429. 

\bibitem{CP94}
S.~Caracciolo and A.~Pelissetto, 
{\it Lattice perturbation theory for $O(N)$-symmetric $\sigma$-models
 with general nearest-neighbour action (I). Conventional perturbation
 theory,}
Nucl.\ Phys.\ B{\bf 420} (1994) 141.

\bibitem{CP95}
S.~Caracciolo and A.~Pelissetto, 
{\it Four-Loop Perturbative Expansion for the Lattice $N$-Vector Model,}
Nucl.\ Phys.\ B{\bf 455} (1995) 619.

\bibitem{Shin}
D.-S.~Shin,
{\it Application of a coordinate space method for the evaluation of
 lattice Feynman diagrams in two dimensions,}
Nucl.\ Phys.\ B{\bf 525} (1998) 457.

\bibitem{Claudia}
S.~Caracciolo, C.~De Grandi and A.~Sportiello, 
{\it Renormalization flow for unrooted forests on a triangular
 lattice,}
Nucl.\ Phys.\ B{\bf 787} (2007) 260.

\bibitem{Ruelle0912}
N.~Azimi-Tafreshi, H.~Dashti-Naserabadi, S.~Moghimi-Araghi and P.~Ruelle,
{\it Abelian Sandpile Model on the Honeycomb Lattice,}
J.\ Stat.\ Mech.\ 
(2010) P02004,
{\tt arXiv:0912.3331}. 

\bibitem{Falcioni-Treves}
M.~Falcioni and A.~Treves, 
{\it The non-linear sigma model: 3-loop renormalization and lattice scaling,}
Nucl.\ Phys.\ B{\bf 265} (1986) 671.

\end{thebibliography}
\end{document}